\documentclass[11pt]{article}
\usepackage{amsmath}
\usepackage{amsfonts}
\usepackage[ansinew]{inputenc}

\newcommand{\EE}{\mathbb{E}}
\newcommand{\PP}{\mathbb{P}}

\newcommand{\VV}{\mathbb{V}}
\newcommand{\ZZ}{\mathbb{Z}}

\newcommand{\corr}{\operatorname{corr}}

\newcommand{\sign}{\operatorname{sign}}
\newcommand{\rk}{\operatorname{rk}}
\newcommand{\Vas}{\operatorname{Vas}}
\newcommand{\med}{\operatorname{med}}
\newcommand{\MAD}{\operatorname{MAD}}
\newcommand{\IMM}{\operatorname{IMM}}
\newcommand{\IM}{\operatorname{IM}}
\newcommand{\DMM}{\operatorname{DMM}}
\newcommand{\KEN}{\operatorname{KEN}}
\newcommand{\MAX}{\operatorname{MAX}}

\begin{document}

\title{Estimation of inter-sector asset correlations}
\author{Christian Meyer\footnote{DZ BANK AG, Platz der Republik, D-60265 Frankfurt. The opinions or recommendations expressed in this article are those of the author and are not representative of DZ BANK AG.} \footnote{E-Mail: {\tt Christian.Meyer@dzbank.de}}}
\date{\today}
\maketitle

\begin{abstract}
Asset correlations are an intuitive and therefore popular way to incorporate event dependence into event risk, e.g., default risk, modeling. In this paper we study the case of estimation of inter-sector asset correlations by separation of cross-sectional dimension and time dimension.
\end{abstract}

\section{Introduction}

In a previous paper \cite{MeyerIntra} we studied estimation of \emph{intra}-sector asset correlations in the context of event risk modelling, remarking that estimation of \emph{inter}-sector asset correlations requires different techniques and would be the subject of a subsequent paper. Now, only 12 years later, that time has come.

The paper is organized as follows. Section \ref{sec_model} specifies the model and the type of panel data from which model parameters are to be estimated. Several estimators for inter-sector asset correlations are proposed in section \ref{sec_estimation}. Section \ref{sec_simstudy} presents the results of an extensive simulation study. Section \ref{sec_conclusion} concludes.

Throughout the paper we will denote the standard normal density by $\varphi(\cdot)$, the standard normal cumulative distribution function by $\Phi(\cdot)$, and the cumulative distribution function of the two-dimensional standard normal distribution with correlation coefficient $\varrho$ by $\Phi_2(\cdot,\cdot,\varrho)$. We will use the symbols $\PP$, $\EE$ and $\VV$ for probabilities, expectation values and variances. For background reading the reader is referred to the precursory paper \cite{MeyerIntra}.


\section{Model setup}
\label{sec_model}

Let $P\sim \Vas(p,\varrho)$ and $\tilde{P}\sim \Vas(\tilde{p},\tilde{\varrho})$ be Vasicek distributed, i.e.,
\begin{align*}
P &= P(Y) = \Phi\left(\frac{\Phi^{-1}(p)-\sqrt{\varrho}\cdot Y}{\sqrt{1-\varrho}}\right),\\
\tilde{P} &= P(\tilde{Y}) = \Phi\left(\frac{\Phi^{-1}(\tilde{p})-\sqrt{\tilde{\varrho}}\cdot \tilde{Y}}{\sqrt{1-\tilde{\varrho}}}\right)
\end{align*}
with standard-normally distributed $Y$, $\tilde{Y}$ and parameters $p,\varrho,\tilde{p},\tilde{\varrho}\in (0,1)$. Let also $\corr(Y,\tilde{Y}) = \gamma$ so that
\[
\begin{pmatrix}
\Phi^{-1}(P)\\\Phi^{-1}(\tilde{P})
\end{pmatrix}
= \begin{pmatrix}
\displaystyle\frac{\Phi^{-1}(p)-\sqrt{\varrho}\cdot Y}{\sqrt{1-\varrho}}\\
\displaystyle\frac{\Phi^{-1}(\tilde{p})-\sqrt{\tilde{\varrho}}\cdot \tilde{Y}}{\sqrt{1-\tilde{\varrho}}}
\end{pmatrix}
\sim N_2\left(
\begin{pmatrix} \mu\\ \tilde{\mu}\end{pmatrix},
\begin{pmatrix}\sigma^2 & \gamma\sigma\tilde{\sigma}\\ \gamma\sigma\tilde{\sigma} & \tilde{\sigma}^2\end{pmatrix}
\right)
\]
with
\[
\mu := \EE\left(\Phi^{-1}(P)\right) = \frac{\Phi^{-1}(p)}{\sqrt{1-\varrho}}, \qquad \tilde{\mu} := \EE\left(\Phi^{-1}(\tilde{P})\right) = \frac{\Phi^{-1}(\tilde{p})}{\sqrt{1-\tilde{\varrho}}},
\]
\[
\sigma^2 := \VV\left(\Phi^{-1}(P)\right) = \frac{\varrho}{1-\varrho}, \qquad \tilde{\sigma}^2 := \VV\left(\Phi^{-1}(\tilde{P})\right) = \frac{\tilde{\varrho}}{1-\tilde{\varrho}},
\]
\[
p_2 := \EE\left(P^2\right) = \Phi_2\left(\Phi^{-1}(p),\Phi^{-1}(p),\varrho\right),
\]
\[
\tilde{p}_2 := \EE\left(\tilde{P}^2\right) = \Phi_2\left(\Phi^{-1}(\tilde{p}),\Phi^{-1}(\tilde{p}),\tilde{\varrho}\right)
\]
and
\[
q := \EE\left(P\cdot\tilde{P}\right) = \Phi_2\left(\Phi^{-1}(p),\Phi^{-1}(\tilde{p}),\gamma\cdot\sqrt{\varrho\tilde{\varrho}}\right).
\]
In this setup, the functions $P(\cdot)$ and $\tilde{P}(\cdot)$ can be described as mixing functions for exchangeable Bernoulli mixture models (cf. \cite{EFM2}, Section 11.2). We think of two populations of individuals with exchangeable binary indicator variables (``events'', e.g., defaults of obligors). Average event probabilities are $p$ and $\tilde{p}$, and dependence between events within populations (``sectors'') is determined by the mixing functions and the \emph{intra-sector asset correlations} $\varrho$ and $\tilde{\varrho}$. The parameter $\gamma$ finally determines dependence between the two sectors. Both $\gamma$ and $\gamma\cdot\sqrt{\varrho\tilde{\varrho}}$ (the latter being interpreted as correlation of abstract asset values) are referred to as \emph{inter-sector asset correlation} in the literature.

The paper \cite{MeyerIntra} has been concerned with estimation of the intra-sector parameters $p$, $p_2$ and $\varrho$, with particular interest in the latter. We will now complete the picture by dealing with estimation of $\gamma$. Extension to more than two sectors is straightforward (with a caveat regarding positive semidefiniteness).

We assume that we are equipped with a panel data set of event histories as follows:
\begin{itemize}
\item There are $T>1$ observation dates (\emph{time dimension}).
\item For observation date $t\in\{1,\ldots,T\}$ there are $n(t)>0$ and $\tilde{n}(t)>0$ individuals (\emph{cross-sectional dimension}).
\item For observation date $t\in\{1,\ldots,T\}$ there are $d(t)\in\{0,\ldots,n(t)\}$ and $\tilde{d}(t)\in\{0,\ldots,\tilde{n}(t)\}$ observations, i.e., individuals for which an event has taken place.
\item The mechanism driving the events for the single observation dates is adequately specified by the above model, with the parameters $p$, $\varrho$, $\tilde{p}$, $\tilde{\varrho}$ and $\gamma$ being constant over time, and the (latent) underlying factors $(Y, \tilde{Y})$ being drawn i.i.d. for different observation dates.
\end{itemize}

As in \cite{MeyerIntra}, the basic strategy for estimation of parameters will be separation of dimensions:
\begin{enumerate}
\item For observation date $t\in\{1,\ldots,T\}$, we will use $d(t)$ and $n(t)$ to estimate the realizations
\[
p(t) \text{ of } P, \qquad \tilde{p}(t) \text{ of } \tilde{P},
\]
\[
g(t) \text{ of } \Phi^{-1}(P), \qquad \tilde{g}(t) \text{ of } \Phi^{-1}(\tilde{P}),
\]
\[
p_2(t) \text{ of } P^2, \qquad \tilde{p}_2(t) \text{ of } \tilde{P}^2,
\]
and $q(t)$ of $P\cdot\tilde{P}$. That is, we will be working in the cross-sectional dimension.
\item We will use these cross-sectional estimates to estimate $\gamma$ (and $p$, $\tilde{p}$, $p_2$, $\tilde{p}_2$, $\varrho$, $\tilde{\varrho}$ where needed). That is, we will be working in the time dimension.
\end{enumerate}

\section{Estimation}
\label{sec_estimation}

\subsection{Indirect moment matching (IMM)}
\label{subsec_estimation_IMM}

Since
\[
\gamma = \corr\left(Y,\tilde{Y}\right) = \corr\left(\Phi^{-1}(P), \Phi^{-1}(\tilde{P})\right)
\]
we have the standard estimator
\[
\gamma^{\IMM} := \frac{\sum_{t=1}^T \left(g(t) - g\right)
\cdot\left(\tilde{g}(t) - \tilde{g}\right)}
{\sqrt{\left(\sum_{t=1}^T \left(g(t) - g\right)^2\right)
\cdot\left(\sum_{t=1}^T \left(\tilde{g}(t) - \tilde{g}\right)^2\right)}}
\]
where
\[
g := \frac{1}{T}\sum_{t=1}^T g(t), \qquad
\tilde{g} := \frac{1}{T}\sum_{t=1}^T \tilde{g}(t).
\]
In the cross-sectional dimension we only need estimates $g(t)$ and $\tilde{g}(t)$. The main problem is dealing with very low or very high numbers $d(t)$, $\tilde{d}(t)$ of observations. As in \cite{MeyerIntra}, in the simulation study we will choose
\[
g(t) := \Phi^{-1}\left(\frac{d(t)+0.6}{n(t)+1.2}\right),\qquad
\tilde{g}(t) := \Phi^{-1}\left(\frac{\tilde{d}(t)+0.6}{\tilde{n}(t)+1.2}\right).\\
\]
We will speak of $\gamma^{\IMM}$ as an \emph{indirect moment matching} estimator (IMM).

Bias will turn out to be an issue for the IMM estimator. We will therefore also try some bias correction as follows:
\begin{itemize}
\item Estimate $\gamma$ (as $\gamma^{\IMM}$), $\rho$, $\tilde{\rho}$, $p$, $\tilde{p}$ from the data.
\item Simulate time series of length $T$ using the estimated parameters and sample sizes $n(t)$, $\tilde{n}(t)$.
\item Estimate $\gamma$ from the time series.
\item Repeat this $M$ times, obtain estimates $\gamma_1,\ldots,\gamma_M$.
\end{itemize}
Finally, estimate $\gamma$ via correcting for the average bias:
\[
\gamma^{\IM 2} := \max\left(-1,\min\left(1,\gamma^{\IMM} + \left(\gamma^{\IMM} - \frac{1}{M}\cdot\sum_{i=1}^M \gamma_i\right)\right)\right)
\]
The result will depend, of course, on the estimators used for $\rho$, $\tilde{\rho}$, $p$, $\tilde{p}$. It seems natural to use the normal variance IMM estimator (cf. \cite{MeyerIntra}, section 4.3.2) for $\varrho$ and $\tilde{\varrho}$.

Repetition of the process provides a 3-step estimator $\gamma^{\IM 3}$. Note that it has to be decided if the estimates for $\rho$, $\tilde{\rho}$, $p$, $\tilde{p}$ are to be corrected for bias as well. We will not do that, and keep the estimates from the previous step to simulate the time series. Note also that in practice repeated application of the procedure will usually converge.

\subsection{Robust estimation using median absolute deviation}
\label{subsec_estimation_MAD}

Since the IMM estimator presented above might suffer from problems with outliers, it is of interest to try a robust correlation estimator instead. For example (cf. \cite{Shevlyakov}), for a bivariate normal vector $(X,Y)$ with correlation $\gamma$, we can define
\[
U := \frac{X-\EE(X)}{\sqrt{2\cdot\VV(X)}} + \frac{Y-\EE(Y)}{\sqrt{2\cdot\VV(Y)}}, \qquad
V := \frac{X-\EE(X)}{\sqrt{2\cdot\VV(X)}} - \frac{Y-\EE(Y)}{\sqrt{2\cdot\VV(Y)}}
\]
and find
\[
\gamma = \frac{\VV(U)-\VV(V)}{\VV(U)+\VV(V)}.
\]
Now we replace expectation value by median, and standard deviation by median absolute distance,
\[
\MAD(X): = \med\left(\left|X-\med(X)\right|\right).
\]
We set
\[
u := \frac{X-\med(X)}{\MAD(X)} + \frac{Y-\med(Y)}{\MAD(Y)}, \qquad
v := \frac{X-\med(X)}{\MAD(X)} - \frac{Y-\med(Y)}{\MAD(Y)}
\]
and estimate
\[
\gamma^{\MAD} := \frac{\MAD^2(u)-\MAD^2(v)}{\MAD^2(u)+\MAD^2(v)}.
\]
Note that the factor $\sqrt{2}$ and the factor between standard deviation and MAD can be ignored. In the simulation study, we will apply this estimator to $(\Phi^{-1}(P), \Phi^{-1}(\tilde{P}))$ as in section \ref{subsec_estimation_IMM}. Median for a sample $x_1\leq \ldots \leq x_{2M}$ will be estimated by $(x_{M}+x_{M+1})/2$. We will speak of $\gamma^{\MAD}$ as a \emph{median absolute distance} estimator (MAD). For alternatives (with possibly superior performance but inferior names) cf. \cite{Shevlyakov}.

\subsection{Direct moment matching (DMM)}
\label{subsec_estimation_DMM}

The idea is to solve the equation
\[
\label{eq_DMM}
q = \Phi_2\left(\Phi^{-1}(p),\Phi^{-1}(\tilde{p}),\delta\right)
\]
for $\delta := \gamma\cdot\sqrt{\varrho\tilde{\varrho}}$, and then extract $\gamma = \delta / \sqrt{\varrho\tilde{\varrho}}$, provided $\varrho$ and $\tilde{\varrho}$ have been estimated before. The equation has a unique solution $\delta=\delta(p,\tilde{p},q)\in(-1,1)$ if $p\in(0,1)$, $\tilde{p}\in(0,1)$ and
\begin{align*}
q & \geq \max(0,p+\tilde{p}-1) = \Phi_2\left(\Phi^{-1}(p),\Phi^{-1}(\tilde{p}),-1\right),\\
q & \leq \min(p,\tilde{p}) = \Phi_2\left(\Phi^{-1}(p),\Phi^{-1}(\tilde{p}),1\right).
\end{align*}
As in \cite{MeyerIntra} we will use the estimators
\[
p^M := \frac{1}{T} \sum_{t=1}^T \frac{d(t)}{n(t)}, \qquad \tilde{p}^M := \frac{1}{T} \sum_{t=1}^T \frac{\tilde{d}(t)}{\tilde{n}(t)}
\]
for $p$ and $\tilde{p}$ (in comparison to the IMM estimator, low numbers $d(t)$, $\tilde{d}(t)$ of observations in the cross-sectional dimension are less of a problem). A natural unbiased estimator for $q$ is given by
\[
q^M := \frac{1}{T} \sum_{t=1}^T \frac{d(t)}{n(t)}\cdot \frac{\tilde{d}(t)}{\tilde{n}(t)}.
\]
Moreover, we have
\begin{align*}
q^M & \geq \max\left(0,\frac{1}{T}\sum_{t=1}^T \left(\frac{d(t)}{n(t)}+\frac{\tilde{d}(t)}{\tilde{n}(t)}-1\right)\right) = \max\left(0,p^M + \tilde{p}^M - 1\right),\\
q^M & \leq \min\left(\frac{1}{T}\sum_{t=1}^T \frac{d(t)}{n(t)},\frac{1}{T}\sum_{t=1}^T \frac{\tilde{d}(t)}{\tilde{n}(t)}\right) = \min\left(p^M,\tilde{p}^M\right).
\end{align*}
Hence it is always possible to solve for $\delta^M := \delta(p^M,\tilde{p}^M,q^M)$. The situation is more subtle for $\gamma$. We propose using the unbiased estimator
\[
p_2^M := \frac{1}{T} \sum_{t=1}^T \frac{d(t)(d(t) - 1)}{n(t)(n(t) - 1)}.
\]
for $p_2$, and then solving for $\varrho^M:= \varrho\left(p^M,p_2^M\right)$ in
\[
\label{eq_DMM_intra}
p_2^M = \Phi_2\left(\Phi^{-1}(p^M),\Phi^{-1}(p^M),\varrho^M\right).
\]
As noted in \cite{MeyerIntra}, we should set $\varrho^M:=0$ if $p_2^M < (p^M)^2$. Finally, we set
\[
\gamma^{\DMM} := \delta^M / \sqrt{p_2^M\cdot \tilde{p}_2^M}
\]
and speak of $\gamma^{\DMM}$ as a \emph{direct moment matching} estimator (DMM).

\subsection{Maximum of IMM and DMM}
\label{subsec_estimation_MAX}

For reasons that will become clear from the discussion of the simulation study, we define another estimator by
\[
\gamma^{\MAX} :=
\begin{cases}
\gamma^{\IMM}, & \qquad\text{if } \left|\gamma^{\IMM}\right| > \left|\gamma^{\DMM}\right|,\\
\gamma^{\DMM}, & \qquad\text{if } \left|\gamma^{\IMM}\right| \leq \left|\gamma^{\DMM}\right|.
\end{cases}
\]

\subsection{Estimation based on measures of concordance}
\label{subsec_estimation_KEN}

For a vector $(X,Y)$ of continuous random variables, measures of concordance (cf. \cite{EFM2}, Section 7.2.3) are scalar measures with the useful property that they are invariant under strictly increasing transformations of the random variables. For example, Kendall's tau is given by
\begin{align*}
\varrho_{\tau}(X,Y) &= \PP\left((X-\hat{X})(Y-\hat{Y}) > 0\right) - \PP\left((X-\hat{X})(Y-\hat{Y}) < 0\right)\\
& = \EE\left(\sign((X-\hat{X})(Y-\hat{Y}))\right)
\end{align*}
where $(\hat{X},\hat{Y})$ is an independent copy of $(X,Y)$. Since the function $\Phi^{-1}(\cdot)$ is strictly increasing, we find
\[
\varrho_{\tau}\left(P,\tilde{P}\right) = \varrho_{\tau}\left(\Phi^{-1}(P),\Phi^{-1}(\tilde{P})\right).
\]
But the vector $\left(\Phi^{-1}(P),\Phi^{-1}(\tilde{P})\right)$ is bivariate normal, and it is known that
\[
\varrho_{\tau}\left(\Phi^{-1}(P),\Phi^{-1}(\tilde{P})\right) = \frac{2}{\pi} \arcsin\left(\corr\left(\Phi^{-1}(P), \Phi^{-1}(\tilde{P})\right)\right) = \frac{2}{\pi} \arcsin\left(\gamma\right).
\]
We can therefore estimate $\gamma$ from Kendall's tau for $(P,\tilde{P})$:
\[
\gamma^{\KEN} := \sin\left(\frac{\pi}{2}\cdot \varrho_{\tau}\left(P,\tilde{P}\right)\right)
\]
Note that there are similar relations for other measures of concordance (cf. \cite{MeyerCopula}, Section 6, for Spearman's rho, Gini's gamma, and Blomqvist's beta). Their usefulness for estimation of $\gamma$ will depend on the information contained. Speaking in Copula terms, Blomqvist's beta only depends on one diagonal, Gini's gamma on the two diagonals, and Kendall's tau and Spearman's rho on the whole square. Therefore, Kendall's tau and Spearman's rho can be expected to perform similarly well, Gini's gamma worse, and Blomqvist's beta worst. In the following we will concentrate on Kendall's tau. For comparison with Spearman's rho, cf. app. \ref{app_spearman}.

The standard estimator for Kendall's tau for two sets $x=\{x_1,\ldots,x_T\}$,  $y=\{y_1,\ldots,y_T\}$ of observations is:
\[
\varrho_{\tau}(x,y) = \frac{2}{T(T-1)}\cdot\sum_{t=1}^T \sum_{s > t} \sign(x_t - x_s)\cdot \sign(y_t - y_s)
\]
However, in the case of the presence of ties or of censoring, where we might have to set some of the signs to zero, it is preferable to also adjust the denominator, i.e. to use the estimator called the tau-b estimator:
\[
\varrho_{\tau}(x,y) = 
\frac{\displaystyle\sum_{t=1}^T \sum_{s > t} \sign(x_t - x_s)\cdot \sign(y_t - y_s)}
{\displaystyle\sqrt{\left(\sum_{t=1}^T \sum_{s > t} \left( \sign(x_t - x_s)\right)^2 \right) \cdot \left(\sum_{t=1}^T \sum_{s > t} \left( \sign(y_t - y_s)\right)^2 \right)}}
\]
In our situation, $x_t=p(t)$ and $x_s=p(s)$. In the simulation study we will simply set $p(t) = d(t) / n(t)$. In order to deal with low numbers of events, it might be useful to play with censoring and bias correction (e.g., as in \cite{NCK}) but we will not explore this further. Note also that we will be using the obvious sign function but there are alternatives, cf. app. \ref{app_sign}.

\section{Simulation study}
\label{sec_simstudy}

Now we will empirically assess the properties of the estimators for $\gamma$ constructed in Section \ref{sec_estimation}. We will choose sets of parameters (9072 sets in total) and generate 10,000 panel data sets for each parameter set. We will evaluate seven estimators for $\gamma$ for each panel data set and each parameter set.

\subsection{Setup}

For each set of parameters and for each panel data set we will use the following estimators for $\gamma$:

\begin{center}
\begin{tabular}{ccc}
\hline
estimator & type & reference\\
\hline
\rule[-2mm]{0mm}{6mm}$\gamma^{\IMM}$ & Indirect moment matching (IMM) & \ref{subsec_estimation_IMM}\\
\rule[-2mm]{0mm}{6mm}$\gamma^{\IM 2}$ & 2-step IMM for bias correction & \ref{subsec_estimation_IMM}\\
\rule[-2mm]{0mm}{6mm}$\gamma^{\IM 3}$ & 3-step IMM for bias correction & \ref{subsec_estimation_IMM}\\
\rule[-2mm]{0mm}{6mm}$\gamma^{\MAD}$ & Median absolute deviation (MAD) & \ref{subsec_estimation_MAD}\\
\rule[-2mm]{0mm}{6mm}$\gamma^{\DMM}$ & Direct moment matching (DMM) & \ref{subsec_estimation_DMM}\\
\rule[-2mm]{0mm}{6mm}$\gamma^{\MAX}$ & Maximum of IMM and DMM & \ref{subsec_estimation_MAX}\\
\rule[-2mm]{0mm}{6mm}$\gamma^{\KEN}$ & (Transformation of) Kendall's tau & \ref{subsec_estimation_KEN}\\
\hline
\end{tabular}
\end{center}
The parameters will be chosen as follows:
\begin{itemize}
\item Time horizon: $T \in\{25,50,100,200,400,800\}$
\item Population size: $n(t)=\tilde{n}(t) := n\in\{100,200,400,800,1600,3200\}$
\item Event probability $p=\tilde{p}\in\{1\%, 2\%, 4\%, 8\%, 16\%, 32\%\}$
\item Intra-sector asset correlation $\varrho=\tilde{\varrho}\in\{1\%, 2\%, 4\%, 8\%, 16\%, 32\%\}$
\item Inter-sector asset correlation $\gamma\in\{-1, -0.5, -0.25, 0, 0.25, 0.5, 1\}$
\end{itemize}
The number of simulations for bias correction (IM2 and IM3 estimators, cf. section \ref{subsec_estimation_IMM}) will be set to $M=100$.

\subsection{Results}

We will discuss some interesting aspects of the results of the simulation study. A complete table listing empirical bias, standard deviation, root mean squared error (RMSE), minimum, 5\% quantile, 25\% quantile, median, 75\% quantile, 95\% quantile, and maximum for each scenario and each estimator is available from the author on request.

Table \ref{table1} presents average standard deviation and RMSE, stratified according to the single parameters. Note that it would make less sense to display average bias (instead of RMSE) since bias is almost symmetrical with respect to $\gamma$. The very last line in the table contains the overall averages.

\begin{table}
\begin{center}
\begin{footnotesize}
\setlength{\arraycolsep}{3pt}
\[
\hspace*{-15mm}
\begin{array}{rrrrrrrrrrrrrrrr}
 & \multicolumn{7}{c}{\text{standard deviation}} &  &  \multicolumn{7}{c}{\text{RMSE}}\\
\cline{2-8}\cline{10-16}
\rule[-1mm]{0mm}{4mm}T & \IMM & \IM2 & \IM3 & \MAD & \DMM & \MAX & \KEN &  & \IMM & \IM2 & \IM3 & \MAD & \DMM & \MAX & \KEN\\
25 & 0.157 & 0.191 & 0.227 & 0.253 & 0.242 & 0.238 & 0.175 &  & 0.230 & 0.222 & 0.240 & 0.324 & 0.256 & 0.250 & 0.236\\
50 & 0.110 & 0.132 & 0.157 & 0.179 & 0.178 & 0.174 & 0.121 &  & 0.191 & 0.169 & 0.172 & 0.253 & 0.186 & 0.182 & 0.190\\
100 & 0.077 & 0.093 & 0.109 & 0.129 & 0.130 & 0.127 & 0.084 &  & 0.166 & 0.135 & 0.127 & 0.209 & 0.135 & 0.132 & 0.161\\
200 & 0.054 & 0.065 & 0.077 & 0.094 & 0.094 & 0.092 & 0.059 &  & 0.150 & 0.113 & 0.097 & 0.179 & 0.097 & 0.095 & 0.143\\
400 & 0.038 & 0.046 & 0.054 & 0.069 & 0.067 & 0.066 & 0.041 &  & 0.140 & 0.098 & 0.078 & 0.160 & 0.070 & 0.069 & 0.131\\
800 & 0.027 & 0.032 & 0.038 & 0.051 & 0.048 & 0.048 & 0.029 &  & 0.134 & 0.088 & 0.065 & 0.148 & 0.050 & 0.050 & 0.123\\
\\
 & \multicolumn{7}{c}{\text{standard deviation}} &  &  \multicolumn{7}{c}{\text{RMSE}}\\
\cline{2-8}\cline{10-16}
\rule[-1mm]{0mm}{4mm}n & \IMM & \IM2 & \IM3 & \MAD & \DMM & \MAX & \KEN &  & \IMM & \IM2 & \IM3 & \MAD & \DMM & \MAX & \KEN\\
100 & 0.088 & 0.123 & 0.167 & 0.156 & 0.209 & 0.207 & 0.102 &  & 0.282 & 0.238 & 0.225 & 0.343 & 0.228 & 0.224 & 0.270\\
200 & 0.084 & 0.109 & 0.136 & 0.141 & 0.160 & 0.158 & 0.093 &  & 0.226 & 0.183 & 0.169 & 0.270 & 0.168 & 0.165 & 0.217\\
400 & 0.079 & 0.096 & 0.112 & 0.131 & 0.124 & 0.121 & 0.086 &  & 0.176 & 0.139 & 0.127 & 0.215 & 0.127 & 0.125 & 0.170\\
800 & 0.074 & 0.085 & 0.094 & 0.121 & 0.100 & 0.098 & 0.080 &  & 0.135 & 0.106 & 0.100 & 0.174 & 0.101 & 0.099 & 0.133\\
1600 & 0.070 & 0.077 & 0.081 & 0.115 & 0.087 & 0.084 & 0.075 &  & 0.105 & 0.085 & 0.084 & 0.145 & 0.088 & 0.086 & 0.106\\
3200 & 0.067 & 0.071 & 0.073 & 0.110 & 0.080 & 0.077 & 0.072 &  & 0.086 & 0.074 & 0.075 & 0.127 & 0.081 & 0.078 & 0.089\\
\\
 & \multicolumn{7}{c}{\text{standard deviation}} &  &  \multicolumn{7}{c}{\text{RMSE}}\\
\cline{2-8}\cline{10-16}
\rule[-1mm]{0mm}{4mm}p & \IMM & \IM2 & \IM3 & \MAD & \DMM & \MAX & \KEN &  & \IMM & \IM2 & \IM3 & \MAD & \DMM & \MAX & \KEN\\
0.01 & 0.086 & 0.117 & 0.154 & 0.150 & 0.190 & 0.187 & 0.099 &  & 0.250 & 0.206 & 0.197 & 0.313 & 0.206 & 0.201 & 0.235\\
0.02 & 0.082 & 0.104 & 0.129 & 0.141 & 0.156 & 0.152 & 0.091 &  & 0.209 & 0.169 & 0.159 & 0.250 & 0.164 & 0.160 & 0.198\\
0.04 & 0.078 & 0.094 & 0.111 & 0.128 & 0.128 & 0.125 & 0.085 &  & 0.173 & 0.140 & 0.131 & 0.210 & 0.133 & 0.130 & 0.168\\
0.08 & 0.075 & 0.087 & 0.098 & 0.122 & 0.108 & 0.106 & 0.080 &  & 0.145 & 0.117 & 0.110 & 0.183 & 0.111 & 0.108 & 0.144\\
0.16 & 0.072 & 0.081 & 0.089 & 0.118 & 0.094 & 0.092 & 0.077 &  & 0.124 & 0.101 & 0.095 & 0.165 & 0.095 & 0.093 & 0.126\\
0.32 & 0.071 & 0.077 & 0.083 & 0.116 & 0.084 & 0.084 & 0.076 &  & 0.110 & 0.091 & 0.087 & 0.153 & 0.085 & 0.084 & 0.114\\
\\
 & \multicolumn{7}{c}{\text{standard deviation}} &  &  \multicolumn{7}{c}{\text{RMSE}}\\
\cline{2-8}\cline{10-16}
\rule[-1mm]{0mm}{4mm}\varrho & \IMM & \IM2 & \IM3 & \MAD & \DMM & \MAX & \KEN &  & \IMM & \IM2 & \IM3 & \MAD & \DMM & \MAX & \KEN\\
0.01 & 0.088 & 0.117 & 0.153 & 0.148 & 0.198 & 0.197 & 0.096 &  & 0.272 & 0.235 & 0.215 & 0.310 & 0.217 & 0.213 & 0.271\\
0.02 & 0.083 & 0.106 & 0.132 & 0.140 & 0.151 & 0.149 & 0.091 &  & 0.218 & 0.180 & 0.162 & 0.258 & 0.159 & 0.157 & 0.215\\
0.04 & 0.078 & 0.095 & 0.112 & 0.132 & 0.117 & 0.116 & 0.086 &  & 0.172 & 0.136 & 0.123 & 0.212 & 0.120 & 0.119 & 0.167\\
0.08 & 0.074 & 0.086 & 0.096 & 0.125 & 0.098 & 0.097 & 0.081 &  & 0.135 & 0.104 & 0.100 & 0.177 & 0.100 & 0.098 & 0.130\\
0.16 & 0.071 & 0.080 & 0.086 & 0.116 & 0.094 & 0.091 & 0.078 &  & 0.111 & 0.086 & 0.091 & 0.155 & 0.095 & 0.092 & 0.106\\
0.32 & 0.069 & 0.076 & 0.083 & 0.114 & 0.101 & 0.096 & 0.077 &  & 0.103 & 0.083 & 0.088 & 0.161 & 0.103 & 0.098 & 0.095\\
\\
 & \multicolumn{7}{c}{\text{standard deviation}} &  &  \multicolumn{7}{c}{\text{RMSE}}\\
\cline{2-8}\cline{10-16}
\rule[-1mm]{0mm}{4mm}\gamma & \IMM & \IM2 & \IM3 & \MAD & \DMM & \MAX & \KEN &  & \IMM & \IM2 & \IM3 & \MAD & \DMM & \MAX & \KEN\\
-1 & 0.037 & 0.046 & 0.041 & 0.068 & 0.059 & 0.054 & 0.041 &  & 0.255 & 0.156 & 0.082 & 0.268 & 0.075 & 0.068 & 0.231\\
-0.5 & 0.085 & 0.103 & 0.126 & 0.142 & 0.140 & 0.138 & 0.094 &  & 0.167 & 0.141 & 0.147 & 0.212 & 0.144 & 0.142 & 0.162\\
-0.25 & 0.096 & 0.117 & 0.144 & 0.157 & 0.160 & 0.158 & 0.106 &  & 0.126 & 0.130 & 0.150 & 0.183 & 0.161 & 0.160 & 0.130\\
0 & 0.101 & 0.122 & 0.150 & 0.163 & 0.168 & 0.172 & 0.110 &  & 0.101 & 0.122 & 0.150 & 0.165 & 0.169 & 0.172 & 0.110\\
0.25 & 0.097 & 0.118 & 0.145 & 0.158 & 0.164 & 0.161 & 0.106 &  & 0.126 & 0.130 & 0.151 & 0.184 & 0.165 & 0.162 & 0.130\\
0.5 & 0.086 & 0.104 & 0.127 & 0.143 & 0.146 & 0.141 & 0.094 &  & 0.164 & 0.140 & 0.150 & 0.214 & 0.149 & 0.143 & 0.161\\
1 & 0.038 & 0.044 & 0.039 & 0.072 & 0.049 & 0.046 & 0.041 &  & 0.241 & 0.143 & 0.079 & 0.259 & 0.064 & 0.060 & 0.224\\
\cline{2-8}\cline{10-16}
\rule[-1mm]{0mm}{4mm}\text{avg} & 0.077 & 0.093 & 0.110 & 0.129 & 0.127 & 0.124 & 0.085 &  & 0.169 & 0.138 & 0.130 & 0.212 & 0.132 & 0.130 & 0.164\\
\end{array}
\]
\end{footnotesize}
\caption{Average standard deviation and RMSE, stratified by the parameters $T$, $n$, $p$, $\varrho$ and $\gamma$}
\label{table1}
\end{center}
\end{table}

As expected, MAD is an inferior version of IMM, reacting similarly to changes in the input parameters but with larger standard deviation and RMSE. Suprisingly, the KEN estimator, although it neglects some information by reducing differences to their signs, is a close competitor to IMM. Note that MAD and KEN might really shine in the case of contaminated data, an issue we will not cover here.

The main battle, however, is between IMM and DMM. In a way, IMM is almost all bias, and DMM is almost all standard deviation. Both estimators profit from larger $T$, larger $n$, larger $p$ and larger $\varrho$ (with caveat: if $\varrho$ becomes very large, the probability of observing zero events increases, which makes estimation more diffcult). Estimation of $\gamma$ is more difficult if $|\gamma|$ is large -- but it is difficult anyway, with standard deviation being an order of magnitude larger than for estimation of intra-correlation $\varrho$ (cf. \cite{MeyerIntra}).

With bias correction of IMM (IM2, IM3) we can trade bias for standard deviation and (kind of) interpolate between IMM and DMM. Note also that IMM and DMM are so consistently different that MAX almost always improves on DMM.

In order to remove the symmetric dependence on $\gamma$ from the picture, table \ref{table2} displays simulation results restricted to $\gamma=0.25$, a value that may be considered relevant in practice. Here it makes sense to display bias instead of RMSE.

\begin{table}
\begin{center}
\begin{footnotesize}
\setlength{\arraycolsep}{3pt}
\[
\hspace*{-20mm}
\begin{array}{rrrrrrrrrrrrrrrr}
 & \multicolumn{7}{c}{\text{standard deviation}} &  &  \multicolumn{7}{c}{\text{bias}}\\
\cline{2-8}\cline{10-16}
\rule[-1mm]{0mm}{4mm}T & \IMM & \IM2 & \IM3 & \MAD & \DMM & \MAX & \KEN &  & \IMM & \IM2 & \IM3 & \MAD & \DMM & \MAX & \KEN\\
25 & 0.197 & 0.241 & 0.297 & 0.304 & 0.312 & 0.313 & 0.218 &  & -0.063 & -0.033 & 0.003 & -0.081 & -0.021 & -0.002 & -0.058\\
50 & 0.138 & 0.167 & 0.206 & 0.219 & 0.232 & 0.228 & 0.152 &  & -0.062 & -0.034 & 0.002 & -0.069 & -0.011 & 0.004 & -0.056\\
100 & 0.097 & 0.117 & 0.144 & 0.159 & 0.169 & 0.164 & 0.106 &  & -0.061 & -0.034 & 0.001 & -0.065 & -0.005 & 0.005 & -0.055\\
200 & 0.069 & 0.082 & 0.101 & 0.117 & 0.123 & 0.118 & 0.074 &  & -0.061 & -0.034 & 0.001 & -0.063 & -0.003 & 0.004 & -0.055\\
400 & 0.048 & 0.058 & 0.071 & 0.086 & 0.088 & 0.085 & 0.052 &  & -0.061 & -0.034 & 0.001 & -0.062 & -0.002 & 0.002 & -0.054\\
800 & 0.034 & 0.041 & 0.050 & 0.063 & 0.062 & 0.061 & 0.037 &  & -0.061 & -0.034 & 0.000 & -0.062 & -0.002 & 0.001 & -0.054\\
\\
 & \multicolumn{7}{c}{\text{standard deviation}} &  &  \multicolumn{7}{c}{\text{bias}}\\
\cline{2-8}\cline{10-16}
\rule[-1mm]{0mm}{4mm}n & \IMM & \IM2 & \IM3 & \MAD & \DMM & \MAX & \KEN &  & \IMM & \IM2 & \IM3 & \MAD & \DMM & \MAX & \KEN\\
100 & 0.100 & 0.140 & 0.198 & 0.170 & 0.245 & 0.245 & 0.115 &  & -0.122 & -0.082 & -0.027 & -0.153 & -0.024 & -0.018 & -0.110\\
200 & 0.099 & 0.129 & 0.169 & 0.160 & 0.199 & 0.197 & 0.110 &  & -0.093 & -0.056 & -0.007 & -0.094 & -0.009 & -0.002 & -0.083\\
400 & 0.098 & 0.119 & 0.147 & 0.157 & 0.162 & 0.159 & 0.106 &  & -0.066 & -0.034 & 0.006 & -0.061 & -0.003 & 0.005 & -0.059\\
800 & 0.097 & 0.111 & 0.130 & 0.155 & 0.138 & 0.134 & 0.103 &  & -0.044 & -0.019 & 0.012 & -0.043 & -0.002 & 0.008 & -0.040\\
1600 & 0.096 & 0.106 & 0.117 & 0.154 & 0.124 & 0.120 & 0.102 &  & -0.027 & -0.009 & 0.013 & -0.030 & -0.003 & 0.010 & -0.025\\
3200 & 0.095 & 0.102 & 0.109 & 0.153 & 0.117 & 0.112 & 0.101 &  & -0.016 & -0.003 & 0.011 & -0.021 & -0.003 & 0.011 & -0.015\\
\\
 & \multicolumn{7}{c}{\text{standard deviation}} &  &  \multicolumn{7}{c}{\text{bias}}\\
\cline{2-8}\cline{10-16}
\rule[-1mm]{0mm}{4mm}p & \IMM & \IM2 & \IM3 & \MAD & \DMM & \MAX & \KEN &  & \IMM & \IM2 & \IM3 & \MAD & \DMM & \MAX & \KEN\\
0.01 & 0.100 & 0.137 & 0.189 & 0.166 & 0.233 & 0.229 & 0.115 &  & -0.105 & -0.064 & -0.009 & -0.134 & -0.021 & -0.009 & -0.091\\
0.02 & 0.098 & 0.126 & 0.163 & 0.164 & 0.197 & 0.192 & 0.110 &  & -0.083 & -0.049 & -0.004 & -0.078 & -0.012 & 0.000 & -0.073\\
0.04 & 0.097 & 0.118 & 0.144 & 0.157 & 0.167 & 0.162 & 0.106 &  & -0.065 & -0.035 & 0.001 & -0.060 & -0.006 & 0.005 & -0.058\\
0.08 & 0.097 & 0.112 & 0.132 & 0.154 & 0.144 & 0.141 & 0.103 &  & -0.049 & -0.025 & 0.005 & -0.051 & -0.003 & 0.007 & -0.045\\
0.16 & 0.096 & 0.108 & 0.123 & 0.154 & 0.128 & 0.126 & 0.102 &  & -0.037 & -0.017 & 0.007 & -0.043 & -0.001 & 0.007 & -0.036\\
0.32 & 0.096 & 0.106 & 0.118 & 0.153 & 0.117 & 0.117 & 0.102 &  & -0.030 & -0.012 & 0.008 & -0.037 & 0.000 & 0.003 & -0.030\\
\\
 & \multicolumn{7}{c}{\text{standard deviation}} &  &  \multicolumn{7}{c}{\text{bias}}\\
\cline{2-8}\cline{10-16}
\rule[-1mm]{0mm}{4mm}\varrho & \IMM & \IM2 & \IM3 & \MAD & \DMM & \MAX & \KEN &  & \IMM & \IM2 & \IM3 & \MAD & \DMM & \MAX & \KEN\\
0.01 & 0.099 & 0.132 & 0.177 & 0.163 & 0.231 & 0.232 & 0.108 &  & -0.117 & -0.085 & -0.043 & -0.119 & -0.017 & -0.014 & -0.114\\
0.02 & 0.098 & 0.125 & 0.161 & 0.163 & 0.185 & 0.185 & 0.107 &  & -0.089 & -0.057 & -0.016 & -0.094 & -0.007 & -0.004 & -0.084\\
0.04 & 0.097 & 0.118 & 0.146 & 0.160 & 0.152 & 0.151 & 0.106 &  & -0.064 & -0.034 & 0.004 & -0.074 & -0.002 & 0.002 & -0.058\\
0.08 & 0.096 & 0.113 & 0.134 & 0.159 & 0.135 & 0.132 & 0.105 &  & -0.044 & -0.017 & 0.017 & -0.056 & -0.002 & 0.006 & -0.038\\
0.16 & 0.096 & 0.110 & 0.126 & 0.152 & 0.134 & 0.128 & 0.105 &  & -0.030 & -0.006 & 0.023 & -0.040 & -0.004 & 0.011 & -0.023\\
0.32 & 0.098 & 0.110 & 0.124 & 0.151 & 0.150 & 0.139 & 0.107 &  & -0.025 & -0.003 & 0.023 & -0.021 & -0.012 & 0.013 & -0.015\\
\cline{2-8}\cline{10-16}
\rule[-1mm]{0mm}{4mm}\text{avg} & 0.097 & 0.118 & 0.145 & 0.158 & 0.164 & 0.161 & 0.106 &  & -0.061 & -0.034 & 0.001 & -0.067 & -0.007 & 0.002 & -0.055\\
\end{array}
\]
\end{footnotesize}
\caption{Average standard deviation and bias for $\gamma=0.25$, stratified by the parameters $T$, $n$, $p$ and $\varrho$}
\label{table2}
\end{center}
\end{table}

Note the interesting pattern for IMM: the parameters $n$, $p$ and $\varrho$ only affect bias, the parameter $T$ only standard deviation. For DMM, on the other hand, bias is negligible but all parameters affect the standard deviation. No surprising effects are visible for the remaining estimators. IM3 seems to slightly over-compensate bias, possibly due to the choice not to bias-correct the estimates for the other parameters (cf. section \ref{subsec_estimation_IMM}).

\section{Conclusion}
\label{sec_conclusion}

Estimation of inter-sector asset correlations is more difficult than estimation of intra-sector asset correlations. Standard deviation and bias of the estimates are an order of magnitude larger. Among the estimators discussed in this paper, the IMM estimator can be recommended. It provides moderate standard deviation and is less sensitive to the other parameters present (event probability, intra-sector asset correlation, sample size). Bias can be corrected via iteration but this will come at the price of increasing standard deviation. If there is suspicion that inter-sector correlation might be close to $\pm 1$, the IMM estimator is useless. However, correlations close to $-1$ seem strange (to me), and correlations close to $+1$ might suggest simplification of the model (via aggregation of systematic factors).

The recommendations from \cite{MeyerIntra} remain in place:
\begin{itemize}
\item Prefer larger number of observation dates to larger event probability.
\item But keep in mind the rule of thumb for event probability (which can be re-stated as: the total number of events observed in a sector should exceed the number $T$ of data points).
\item Think about event definition (maybe include not only defaults but also downgrades), which will also help in ensuring exchangeability.
\end{itemize}
In addition, pay attention to quantification of bias and of estimation uncertainty (cf. \cite{FreiWunsch}, \cite{MeyerModelRisk}, \cite{PNFR}). I also commend the approach taken in \cite{PNFR} where several estimators (including maximum likelihood estimators, which we have not discussed) have been made available as R implementation.


\begin{appendix}

\section{Spearman's rho}
\label{app_spearman}

Spearman's rho for a random vector $(X,Y)$ is given by
\[
\varrho_S(X,Y) = \corr(F(X),G(Y))
\]
where $F$ and $G$ are the marginal distribution functions of $X$ and $Y$. The standard estimator for Spearman's rho uses the empirical distribution function or, equivalently, the ranks of the observations in the ordered samples. Let observations $x=\{x_1,\ldots,x_T\}$, $y=\{y_1,\ldots,y_T\}$ be given. The rank of $x_t$ is defined as
\[
\rk(x_t) := \frac{T+1}{2}+\frac{1}{2}\sum_{s\neq t} \sign(x_t-x_s),
\]
with the convention $\sign(0)=0$. Using
\[
\overline{\rk}(x) := \frac{1}{T} \sum_{t=1}^T \rk(x_t) = \frac{T+1}{2}
\]
the standard estimator reads:
\begin{align*}
\varrho_S(x,y) & = \frac{\sum_{t=1}^T \left(\rk(x_t)-\overline{\rk}(x)\right)\left(\rk(y_t)-\overline{\rk}(y)\right)}
{\sqrt{\left(\sum_{t=1}^T\left(\rk(x_t)-\overline{\rk}(x)\right)^2\right) \cdot \left(\sum_{t=1}^T\left(\rk(y_t)-\overline{\rk}(y)\right)^2\right)}}\\
& = \frac{\displaystyle\sum_{t=1}^T\left(\sum_{s > t} \sign(x_t-x_s)\right)\left(\sum_{s > t} \sign(y_t-y_s)\right)}
{\displaystyle\sqrt{\left(\sum_{t=1}^T\left(\sum_{s > t} \sign(x_t-x_s)\right)^2\right) \cdot \left(\sum_{t=1}^T\left(\sum_{s > t} \sign(y_t-y_s)\right)^2\right)}}
\end{align*}
It becomes apparent that Spearman's rho can be expected to perform similarly to Kendall's tau. In the setting of section \ref{subsec_estimation_KEN}, we might use the relationship
\[
\varrho_S\left(P,\tilde{P}\right) = \varrho_S\left(\Phi^{-1}(P),\Phi^{-1}(\tilde{P})\right)
= \frac{6}{\pi} \arcsin\left(\frac{\gamma}{2}\right)
\]
and hence estimate:
\[
\gamma = 2\cdot\sin\left(\frac{\pi}{6}\cdot \varrho_S\left(P,\tilde{P}\right)\right)
\]

\section{Alternative sign function}
\label{app_sign}

If $x_t$, $x_s$ are given as distributions instead of fixed values (e.g., because they have been estimated the Bayesian way), we can define:
\[
\sign(x_t-x_s) := \PP(x_t > x_s)-\PP(x_t < x_s).
\]
Furthermore, if the distributions are continuous then we can write
\[
\PP(x_t > x_s) - \PP(x_t < x_s) = 1 - 2\cdot\PP(x_t \leq x_s).
\]

As an example, consider Bayesian estimation of binomial proportions with uniform (i.e., $\beta(1,1)$) prior: let $X_1\sim\beta(d_1 + 1,n_1 - d_1 + 1)$, $X_2\sim\beta(d_2 + 1,n_2 - d_2 + 1)$ be independent beta-distributed random variables with $d_1,n_1,d_2,n_2\in\ZZ$, $0\leq d_1\leq n_1$, $0\leq d_2\leq n_2$. We apply the convolution formula, then \cite{AS}, 26.5.7 and some manipulation of binomial coefficients to obtain
\begin{align*}
\PP(X_2 \leq X_1) & = \PP(X_2-X_1\leq 0)\\
& = \frac{\int_0^1 (1-I_t(d_1+1,n_1-d_1+1))\cdot t^{d_2}\cdot (1-t)^{n_2-d_2} \,dt}{B(d_2+1,n_2-d_2+1)}\\
& = \frac{\int_0^1 \sum_{i=0}^{d_1} \binom{n_1+1}{i}\cdot\left(\frac{t}{1-t}\right)^i \cdot t^{d_2}\cdot(1-t)^{n_1+n_2+1-d_2} \,dt}{B(d_2+1,n_2-d_2+1)}\\
& = \frac{\sum_{i=0}^{d_1} \binom{n_1+1}{i}\int_0^1 t^{d_2+i}\cdot(1-t)^{n_1+n_2+1-d_2-i}\,dt}{B(d_2+1,n_2-d_2+1)}\\
& = \frac{\sum_{i=0}^{d_1} \binom{n_1+1}{i} B(d_2+i+1,n_1+n_2+2-d_2-i)}{B(d_2+1,n_2-d_2+1)}\\
& = \binom{n_1+n_2+2}{n_1+1}^{-1} \sum_{i=0}^{d_1}\binom{d_2+i}{d_2}\binom{n_1+n_2+1-d_2-i}{n_2-d_2}.
\end{align*}

\end{appendix}

\end{document}